# Ion-exchange synthesis and superconductivity at 8.6 K of $Na_2Cr_3As_3$ with quasi-one-dimensional crystal structure

Qing-Ge Mu[†,‡], Bin-Bin Ruan[†,‡], Bo-Jin Pan[†,‡], Tong Liu[†,‡], Jia Yu[†,‡], Kang Zhao[†,‡], Gen-Fu Chen[†,‡,§], and Zhi-An Ren[*,†,‡,§]

[†]Institute of Physics and Beijing National Laboratory for Condensed Matter Physics, Chinese Academy of Sciences, Beijing 100190, China

[‡]School of Physical Sciences, University of Chinese Academy of Sciences, Beijing 100049, China

[§]Collaborative Innovation Center of Quantum Matter, Beijing 100190, China

**Corresponding Author**

* renzhian@iphy.ac.cn

**ABSTRACT:**

A new Cr-based quasi-one-dimensional superconductor $Na_2Cr_3As_3$ was synthesized by an ion-exchange method in sodium naphthalenide solution. The crystals are thread-like and the structure was analyzed by X-ray diffraction with a noncentrosymmetric hexagonal space group $P$-$6m$2 (No. 187), in which the $(Cr_3As_3)^{2-}$ linear chains are separated by $Na^+$ ions, and the refined lattice parameters are $a$ = 9.239(2) Å and $c$ = 4.209(6) Å. The measurements for electrical resistivity, magnetic susceptibility, and heat capacity reveal a superconducting transition with unconventional characteristic at the $T_c$ of 8.6 K, which exceeds that of all previously reported Cr-based superconductors.

The discoveries of Cu-based and Fe-based high-$T_c$ superconductors have vitalized extensive research interests in the unconventional superconductivity associated with magnetic instability especially among 3d transition metal compounds during the last 30 years.[1-3] Recently, the early transition metal Cr-based binary compound CrAs was found to undergo a superconducting transition at 2 K on the border of a helical antiferromagnetic order upon applying external pressure,[4-5] which is significantly different from the several previously reported Cr-containing alloy superconductors.[6-11] Soon after, superconductivity was discovered at ambient pressure in new types of Cr-based ternary compounds $A_2Cr_3As_3$ and $ACr_3As_3$ (A = K, Rb or Cs) which show very particular quasi-one-dimensional (Q1D) crystal structure and evidences for fascinating spin-triplet electron paring.[12-16] This Cr-based superconducting family has attracted intense interests and is providing a new platform for the understanding of unconventional superconductivity.

For the exploration of new superconductors, chemical pressure effect has been an effective tool to tune the crystal structure and the electronic properties for obtaining new superconductors or achieving higher $T_c$, which is usually realized by substituting part of the structural motif or ions with equivalent ones but different spacial sizes, and it causes a relative change of crystal lattice analogous to external physical pressure effect. This is typically illustrated in the iron-oxypnictide superconductors, as the superconducting $T_c$ dramatically increases from 26 K of $LaO_{1-x}F_xFeAs$ to 55 K of $SmO_{1-x}F_xFeAs$ by the substitution of $[La_2O_2]^{2+}$ with smaller $[Sm_2O_2]^{2+}$.[3, 17-18] For the new superconductors $A_2Cr_3As_3$ (A = K, Rb, Cs), the crystal structure can be regarded as Q1D infinite $(Cr_3As_3)^{2-}$ linear chains which are separated by columns of alkali metal cations $A^+$ within a noncentrosymmetric hexagonal crystal lattice with the space group $P\text{-}6m2$ (No. 187).[12-14] The $(Cr_3As_3)^{2-}$ chain structure is regarded as the key unit to be responsible for the occurrence of superconductivity.[19-21] It was proved that replacing $K^+$ with larger $Rb^+$ or $Cs^+$ in $K_2Cr_3As_3$ results in decreasing $T_c$ from 6.1 K to 4.8 K and 2.2 K, which indicates a positive chemical pressure effect in the $A_2Cr_3As_3$ superconductors.[12-14, 22] But contrarily, the superconducting $T_c$ of $K_2Cr_3As_3$ decreases monotonically under external physical pressures revealed by several reports, which demonstrates a negative physical pressure effect on $T_c$.[23-25] This discrepancy is possibly originated from the unique Q1D anisotropic crystal structure which has different electronic responses on the chain structure and inter-chain coupling to the chemical pressure and physical pressure, respectively.[25] Consequently, the only way to achieve higher $T_c$ in this system might be replacing $A^+$ with much smaller monovalent ions of $Na^+$ or $Li^+$, but all attempts to synthesize $Na_2Cr_3As_3$ or $Li_2Cr_3As_3$ by conventional high-temperature solid-state reaction method or single crystal growth were failed in the past 3 years.[26]

In this letter, we report the successful synthesis of the meta-stable Q1D $Na_2Cr_3As_3$ compound using a soft chemistry method. The $Na_2Cr_3As_3$ is obtained by the spontaneous ion-exchange reaction that has been more widely utilized for layered materials previously.[27-29] And we find superconductivity in this new Cr-based compound by electrical resistivity, magnetic susceptibility and heat capacity characterizations with a much enhanced $T_c$ of 8.6 K.

The single crystals of $Na_2Cr_3As_3$ were prepared by ion-exchange method with sodium naphthalene solution (Naph.-Na) in tetrahydrofuran (THF) using $K_2Cr_3As_3$ crystals as the precursor. At first the high-quality $K_2Cr_3As_3$ crystals were grown by a high-temperature solution growth method as previously reported (see Supporting Information).[12] A 0.1 M Naph.-Na was prepared by dissolving an equimolar amount of naphthalene and sodium into THF which was dehydrated with molecular sieves (4A). Then, 10 mg as grown rod-like $K_2Cr_3As_3$ crystals were immersed into 7.5 mL Naph.-Na in a Teflon liner and loaded into an autoclave, and the Naph.-Na was excess to ensure the ion-exchange process was sufficient with K:Na molar ratio close to 1:17. The autoclave was tightly sealed, and heated in a muffle furnace at 373 K for 50-100 h. After cooling down to room temperature, the obtained samples were washed with THF thoroughly for several times. All sample preparation procedures were carried out in a glove box ($O_2$, $H_2O$ < 0.1 ppm) filled with high-purity Ar gas.

The obtained $Na_2Cr_3As_3$ crystals are extremely unstable in air, and any exposure to air should be avoided when performing measurements on them. The crystal structure was characterized at room temperature by both powder x-ray diffraction (PXRD) using a PAN-analytical x-ray diffractometer and single-crystal x-ray diffraction (SXRD) using a Bruker single crystal x-ray diffractometer with Cu-$K_\alpha$ radiation. The chemical composition was analyzed with both Phenom scanning electron microscopy (SEM) with an energy-dispersive spectrometer (EDS) and inductively coupled plasma-atomic emission spectroscopy (ICP). The resistivity and heat capacity were measured with a Quantum Design physical property measurement system by the standard four-probe method and relaxation method, respectively. The dc magnetic susceptibility measurement was performed with a Quantum Design magnetic property measurement system under zero-field-cooling (ZFC) and field-cooling (FC) modes from 2 K to 10 K under a magnetic field of 10 Oe.

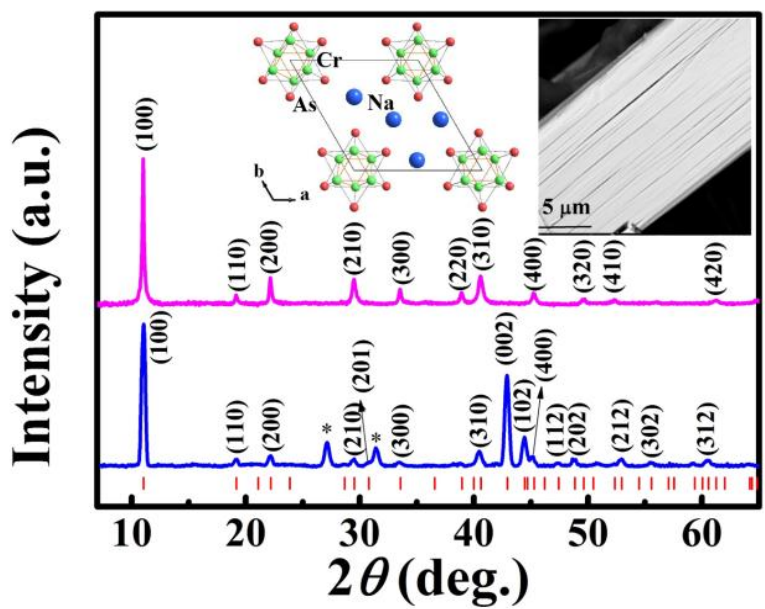

**Figure 1.** The XRD patterns for $Na_2Cr_3As_3$ samples collected by (a) PXRD and (b) SXRD. The left inset is the hexagonal lattice structure of $Na_2Cr_3As_3$, and the right inset shows the SEM micrograph of $Na_2Cr_3As_3$ crystals.

The XRD patterns are shown in Figure 1. Due to the Q1D crystal morphology of the sample, the grains are highly oriented after powderized, hence the PXRD data only show the peaks of ($hk$0) without the information associated with $c$-axis. On the other hand, the patterns collected by SXRD on a bundle of crystals show more complete diffraction peaks for the lattice planes. We note that our attempt to resolve the crystal structure by SXRD on a single piece of crystal all failed due to the split of all crystals during the ion exchange procedure. The XRD patterns are both well indexed with the hexagonal lattice structure with space group $P$-$6m2$ (No. 187) as that of $K_2Cr_3As_3$, except two impurity peaks marked with asterisks.[12, 23] The impurity is identified to be $Cr_2O_5$ which is possibly induced by oxidation during the experimental process due to the high activity of the $Na_2Cr_3As_3$ phase. The crystal lattice structure is also shown as the inset in Figure 1. From the XRD patterns, the lattice parameters were refined to be $a = 9.239(2)$ Å and $c = 4.209(6)$ Å. The corresponding theoretical calculations for Bragg peaks are indicated by the vertical bars for a clear view. Comparing with $K_2Cr_3As_3$,[12, 23] the inter-chain distance shrinks obviously along $a$-axis (about 7.6% from the $K_2Cr_3As_3$ precursor), while the Q1D chain structure shrinks very slightly along $c$-axis (~ 0.5%), which causes a primary chemical pressure perpendicular to the linear $(Cr_3As_3)^{2-}$ chain structure by the replacement of $K^+$ with much smaller $Na^+$. The morphology of the Q1D thread-like crystals is shown by the SEM image as the inset in Figure 1. The atomic ratio of Na:Cr:As measured by EDS is close to 2:3:3, which is also confirmed by the results of ICP measurements, and this verifies the chemical composition of the $Na_2Cr_3As_3$ crystals (see Supporting Information).

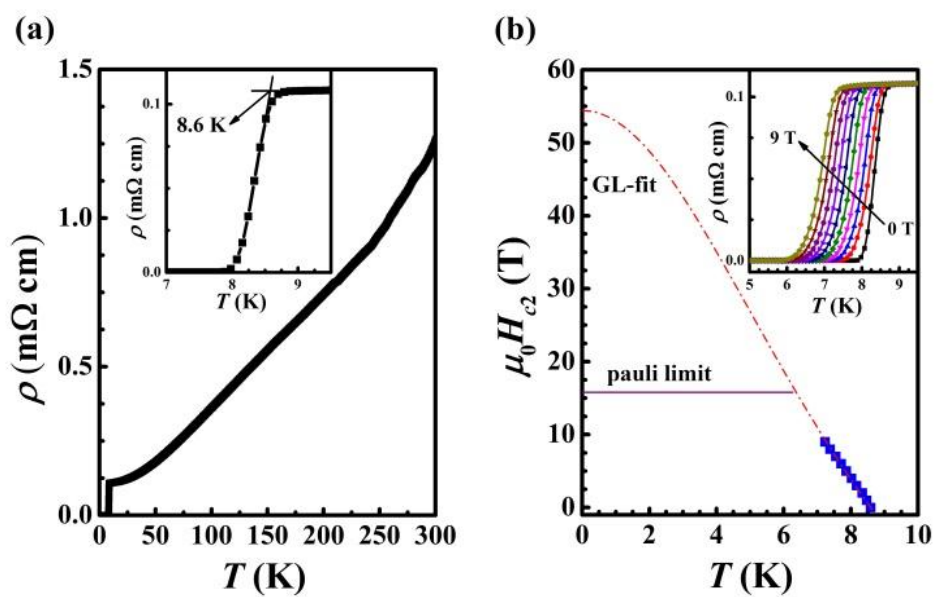

**Figure 2.** (a) The temperature dependence of resistivity for $Na_2Cr_3As_3$ crystals, and the inset exhibits the enlarged view for the superconducting transition. (b) The derived upper critical field with GL-theory and the Pauli paramagnetic limit for $Na_2Cr_3As_3$ crystals, and the inset represents the superconducting transition under different magnetic fields from 0 T to 9 T.

The temperature dependence of resistivity for $Na_2Cr_3As_3$ crystal is characterized from 1.8 K to 300 K at zero field with the electrical current along the $c$-axis as illustrated in Figure 2a, which exhibits metallic behavior in the normal state, and a sudden superconducting transition is observed at low temperature. Notably, the onset superconducting $T_c$ is 8.6 K as shown in the inset in Figure 2a, which is distinctly enhanced by 41% than that of the $K_2Cr_3As_3$ precursor (~ 6.1 K) and being the highest $T_c$ in all previously reported Cr-based superconductors.[4-5, 12-16] The width of superconducting transition is less than 1 K. The room temperature resistivity is about 1.3 mΩ cm, and the residual resistivity ratio (RRR) is about 12, which is much lower than that of $K_2Cr_3As_3$ single crystal (RRR ~ 60).[15, 23] The lower RRR value is possibly due to the crystal defects or lattice deformation induced during the ion-exchange process. In the normal state the temperature dependence of resistivity deviates from linear behavior, which was observed in polycrystals previously but not in single crystals of $K_2Cr_3As_3$.[12, 15, 23] To further characterize the superconductivity under magnetic field, we measured the temperature dependence of resistivity under different magnetic fields from 0 T to 9 T to study the upper critical field ($\mu_0H_{c2}$) with the field perpendicular to the $c$-axis and the electrical current along the $c$-axis. The temperature dependence of resistivity at low temperature is shown in the inset in Figure 2b. Upon applying magnetic field, superconducting transition shifts to lower temperature slowly. The onset superconducting $T_c$ under different magnetic fields is determined the same way as illustrated in the inset in Figure 2a, and the temperature dependence of $\mu_0H_{c2}$ is depicted in Figure 2b. The $\mu_0H_{c2}(T)$ data are fitted with the Ginzburg-Landau theory, $\mu_0H_{c2}(T) = \mu_0H_{c2}(0)(1 - t^2)/(1 + t^2)$, here $t = T/T_c$, and the zero-temperature upper critical field $\mu_0H_{c2}(0)$ is estimated to be 54 T. Considering Pauli paramagnetic limited $\mu_0H_p = 1.84\ T_c \approx 16$ T,[30] the extrapolated $\mu_0H_{c2}(0)$ is much higher than the Pauli paramagnetic limit for weak coupling conventional BCS superconductor. This gives a strong indication for the unconventional superconductivity in $Na_2Cr_3As_3$, which is also observed previously in $A_2Cr_3As_3$,[12-14] $ACr_3As_3$,[15-16] $Li_{0.9}Mo_6O_{17}$[31] and $Sr_2RuO_4$[32] superconductors.

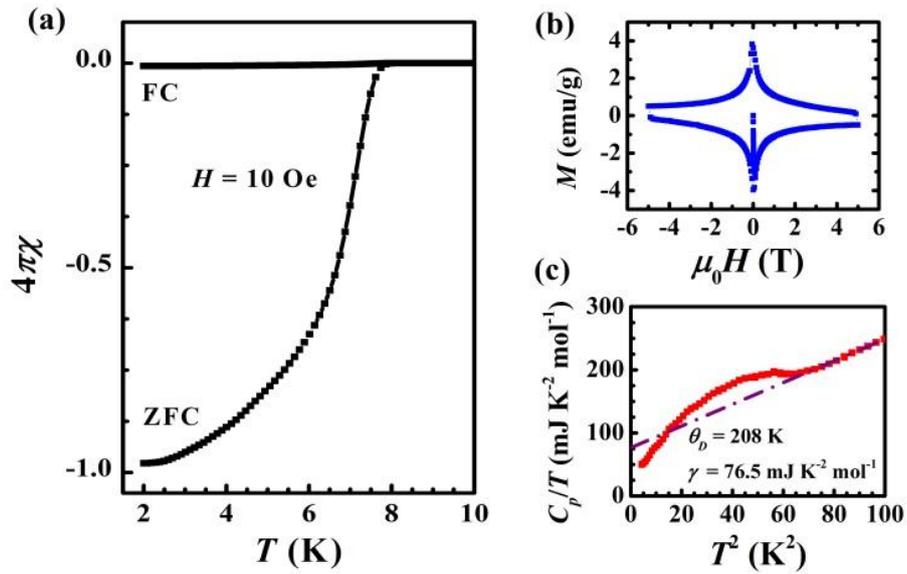

**Figure 3.** (a) The temperature dependence of magnetic susceptibility for $Na_2Cr_3As_3$ crystals. (b) Isothermal magnetization at 2 K from -5 T to 5 T. (c) The temperature dependence of heat capacity depicted as $C_p/T$ vs $T^2$ with linear fitting in the normal state.

To further confirm the superconductivity in $Na_2Cr_3As_3$, the temperature dependence of dc magnetization and heat capacity for the crystals were characterized and shown in Figure 3. The dc magnetization data of $Na_2Cr_3As_3$ were collected under a magnetic field of 10 Oe which was perpendicular to the $c$-axis. Both FC and ZFC data show the diamagnetic transition below an onset $T_c$ of 8.1 K as shown in Figure 3a, which is consistent with the results of resistivity measurement. The shielding volume fraction from ZFC data is nearly 100% at 2 K, suggesting bulk superconductivity in $Na_2Cr_3As_3$, while the much smaller value of FC susceptibility data indicates strong magnetic flux pinning effect caused by crystal defects. Figure 3b shows the isothermal magnetization data at 2 K with magnetic field from -5 T to 5 T, and it reveals the behavior of a typical type-II superconductor for $Na_2Cr_3As_3$.

The temperature dependence of heat capacity is depicted as a relationship of $C_p/T$ vs. $T^2$, which is shown in Figure 3c. The heat capacity of a common metal can be linearly fitted with $C_p/T = \gamma + \beta T^2$, which contains the contributions from electrons and phonons. When fitting the data with the above formula in the normal state, we obtained the Sommerfeld coefficient $\gamma$ as 76.5 mJ K$^{-2}$ mol$^{-1}$, and Debye temperature $\theta_D$ as 208 K calculated from $\theta_D = [(12/5)NR\pi^4/\beta]^{1/3}$. Comparing with $K_2Cr_3As_3$, the close values of large $\gamma$ indicate similar strong electron correlations in $Na_2Cr_3As_3$.[12] The clear heat capacity anomaly is observed at about 8 K, suggesting the occurrence of a superconducting phase transition. This result further confirms the superconductivity observed in resistivity and magnetization measurements. However, the smaller value of electronic heat capacity jump $\Delta C_p/\gamma T_c$ (~ 0.69) and the wider transition width compared with that of $K_2Cr_3As_3$ indicates the low sample quality with amounts of defects produced during the ion exchange process.[12]

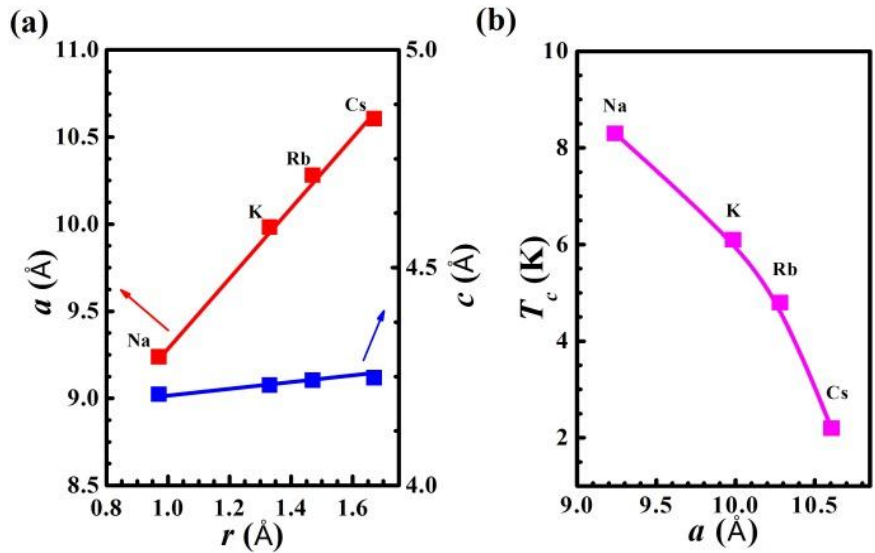

**Figure 4.** (a) The variation of lattice parameters with the change of A$^+$ ionic radius in $A_2Cr_3As_3$. (b) Superconducting $T_c$ changes with the lattice parameter $a$.

The lattice parameters upon the change of A$^+$ ionic radius for all the existing $A_2Cr_3As_3$ (A = Na, K, Rb or Cs) superconductors are summarized in Figure 4a.[12-14] It shows that when decreasing the cation radius from Cs$^+$ to Na$^+$, the hexagonal crystal lattice shrinks very little along the Q1D $(Cr_3As_3)^{2-}$ chain direction (the $c$-axis), while it shrinks obviously along the $a$-axis, which shows enhanced inter-chain interactions and large chemical pressure on the radial direction of the chain structure. This makes it much different with the hydrostatic physical pressure effect. Correspondingly, as shown in Figure 4b, the superconducting $T_c$ increases dramatically from 2.2 K to 8.6 K (about 4 times enhancement) by the replacement of alkali metal cation from Cs$^+$ to Na$^+$, which shows very large chemical pressure effect in this Cr-233 superconducting family, and it indicates the strong correlations between superconductivity and the inter-chain interactions. Hence it might be feasible to increase the $T_c$ in this family by further reducing the inter-chain distance through chemical replacement or possible uniform radial physical pressure.

In summary, we synthesized a new Cr-based Q1D ternary compound $Na_2Cr_3As_3$ through the ion-exchange method, and superconductivity was discovered in this compound with the transition temperature of 8.6 K, which is the highest $T_c$ in all Cr-based superconductors. The characterizations of resistivity and heat capacity revealed unconventional superconductivity with strong electronic correlations in $Na_2Cr_3As_3$. The dramatic enhancement of $T_c$ in this $A_2Cr_3As_3$ series indicates large chemical pressure effect on superconductivity.

## ACKNOWLEDGMENT

The authors are grateful for the financial supports from the National Natural Science Foundation of China (No. 11474339 and 11774402), the National Basic Research Program of China (973 Program, No. 2016YFA0300301) and the Youth Innovation Promotion Association of the Chinese Academy of Sciences.